\documentclass[a4paper]{jpconf}
\usepackage{graphicx}
\usepackage{cite}
\usepackage{gensymb}
\usepackage{amsmath}
\usepackage{amssymb}

\newcommand{\rr}{\textbf{r}}

\DeclareRobustCommand{\&}{%
  \ifdim\fontdimen1\font>0pt
    \textsl{\symbol{`\&}}%
  \else
    \symbol{`\&}%
  \fi
}

\begin{document}
\title{Origin of hardening and universality of cosmic rays spectra in GV--PV rigidity region}

\author{A A Lagutin, N V Volkov, R I Raikin and A G Tyumentsev}

\address{Altai State University, Radiophysics and Theoretical Physics Department. 61 Lenin ave., Barnaul, 656049, Russia.}

\ead{lagutin@theory.asu.ru, volkov@theory.asu.ru, altraikin@gmail.com, tyumentsev@theory.asu.ru}

\begin{abstract}
Recent balloon-borne and satellite experiments have established new features in the behavior of the spectra of cosmic rays. An analysis of all the data showed that hardening of most abundant primary cosmic ray nuclei spectra with increasing rigidity is observed in $>200$~GV region. At the same time, rigidity dependences of secondary cosmic rays are distinctly different. The AMS-02 data show that above 200 GV the secondary cosmic rays Li, Be, B  harden more than the primary He, C, O. 

In this paper we discuss a new scenario that self-consistently describes these new features of the  cosmic ray spectra. We demonstrate that the measured by the AMS-02 changes in the slope   of the cosmic ray spectra caused by the transition from the contribution of multiple distant Galactic sources, including  the nuclear interactions of the particles accelerated by these sources with an interstellar medium during their wandering in the Galaxy, to the contribution of mainly local ones. We also found that the spectral universality is observed at rigidity $R> 10^5$~GV.
\end{abstract}

\section{Introduction}

\begin{figure}

\begin{center}
\includegraphics[width=.9\textwidth]{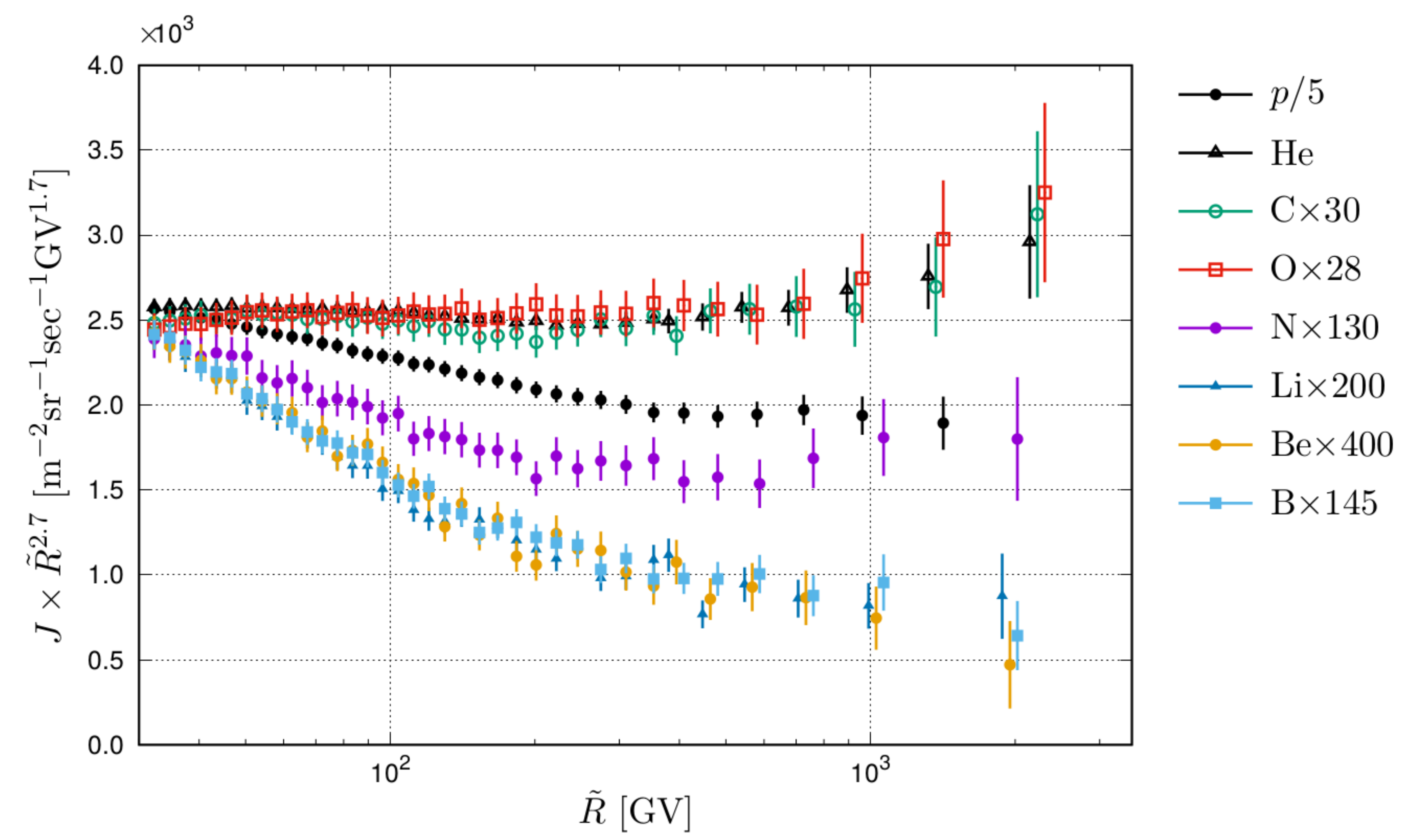}
\end{center}
\caption{The AMS-02 primary and secondary cosmic ray fluxes with their total errors multiplied by $\tilde{R}^{2.7}$ as a function of rigidity above 30 GV~\cite{Aguilar:2015P,Aguilar:2015,Aguilar:2017CO,Aguilar:2018Li,Aguilar:2018N}. The fluxes were rescaled as indicated. The He, O, Li, and B data points above 400 GV are displaced horizontally  by $\sim$ 2\%}\label{fig:universality-spectra} 
\end{figure}

In the last decade, measurements of the cosmic ray (CR) nuclei in the GV--TV rigidity region by new-generation balloon-borne and satellite instruments~\cite{Panov:2009,Ahn:2010,Adriani:2011, Aguilar:2015P,Aguilar:2015,Aguilar:2017CO,Aguilar:2018Li,Aguilar:2018N} have established new features in behavior of their spectra. It was found that both the spectra of most abundant primary CR nuclei proton, He, C, O, N, and the secondary cosmic rays  Li, Be, B  at rigidity $R> 100-200$~GV exhibit a hardening with increasing rigidity. As follows from figure~\ref{fig:universality-spectra}, they deviate from a single power law. Above 60 GV, the three fluxes He, C, O measured by the AMS-02 have identical rigidity dependence within the measurement errors and the ratios He/O and C/O are well fit by a constant value~\cite{Aguilar:2017CO}. 

The rigidity dependences of secondary and primary CRs are distinctly different. The AMS-02 data show that above 200 GV the secondary cosmic rays harden more than the primary He, C, O~\cite{Aguilar:2018Li}.

These newly discovered features are not easy to explain under standard scenario
of cosmic ray origin, acceleration and their propagation in the Galaxy. Under the standard theory,  
the primary nuclei are thought to be produced, at least to several PV, by supernova remnant shock waves by diffusive shock acceleration mechanism that predicts power-law spectra $J\propto R^{-\gamma}$ with slope $\gamma\approx 2.0-2.2$. The subsequeant CR transport in the turbulent galactic magnetic fields is modeled as a diffusion process in quasi-homogeneous medium  with the diffusion coefficient $D(R)=D_0{(R/1~\text{GV})}^{\delta}$, with $\delta \approx (0.3 - 0.8)$. 
Under these assumptions, the spectrum of primary nucleus $i$ generated by the global-scale steady state distribution of sources $S(\rr)$ is described by a single power law with index $\eta = \gamma + \delta$, which is clearly at odds with the observed hardening of CR hadrons at GV--TV region.

Due to the fact that the predicted by the standard scenario observed spectrum is determined by the parameters $\gamma$, $\delta$  and the distribution of   sources  $S(\rr)$,
the proposed solutions to the hardening problem include acceleration mechanisms modification, change of the CR diffusion properties in the different regions of the Galaxy and superposition of local and distant sources (see, e.g.,~\cite{Biermann:2010,Erlykin:2012,Thoudam:2012,Tomassetti:2012,Vladimirov:2012,Blasi:2012,Ptuskin:2013,Bernard:2013,Erlykin:2015,Zatsepin:2006}).

In this paper, we continue to investigate the spectral hardening problem observed at GV--TV rigidity region  within the framework of the approach we proposed in our paper~\cite{Lagutin:2001}. In that paper, we point to possible anomalies in the spectra of protons and helium, due to differences in the behavior of the energy spectra of distant and nearby sources in a highly inhomogeneous galactic medium. However, the lack of experimental data for the range of $\sim 200-3000$ GeV at that time did not allow  us to discuss these new features in~\cite{Lagutin:2001}.

In the current study, using the precision measurement of the primary and secondary cosmic ray nuclei  spectra in the AMS-02  experiment~\cite{Aguilar:2015P,Aguilar:2015,Aguilar:2017CO,Aguilar:2018Li,Aguilar:2018N}, we demonstrate that the observed changes in the slope   of these spectra 
caused by the transition from the contribution of multiple distant Galactic sources, including  nuclear interactions of particles accelerated by these sources with an interstellar medium during their wandering in the Galaxy, to the contribution of mainly local ones.

\section{Model}

The following basic principles and key assumptions were included in our model.
\begin{enumerate}

\item All particles with a rigidity $30  \lesssim R \lesssim  5 \cdot 10^{7}$ GV, observed in the Solar system and on the Earth, are accelerated mainly by Galactic sources. The spectrum of accelerated particles in the sources is described by the power law  $J\propto R^{-\gamma}$. 

\item CR sources are divided into two groups as in our paper~\cite{Lagutin:2001}. The first group includes the multiple old distant ($r\geq 1$~kpc) sources, the second group consists of nearby ($r< 1$~kpc) young ($t< 10^6$~yr) sources. The spatial distribution of sources also suggests the separation of the observed CR flux of the nucleus $i$  into two components as follows:
$$
J(\rr,t,R) = J_G(\rr,R)+ J_L(\rr,t,R).
$$
Here
\begin{itemize}
\item $J_G$ is the global spectrum component determined by the multiple old ($t\geq 10^6$~yr) distant ($r\geq 1$~kpc) sources, including  the contribution of nuclear interactions of the particles accelerated by these steady state sources with an interstellar medium during their wandering in the Galaxy.
\item $J_L$ is the local component, i.e. the contribution of nearby ($r< 1$~kpc) young ($t< 10^6$~yr) sources; the effect of nuclear interactions of particles accelerated in these sources with an interstellar medium on the observed spectrum  is negligible. It is not taken into account in this work. 
\end{itemize}
\item The highly inhomogeneous distribution of matter and magnetic fields in the Galaxy  leads to the anomalous diffusion of CRs~\cite{Lagutin:2017V}. Anomalous diffusion is manifested, in particular, by abnormally large free paths of particles (so-called ``L\'{e}vy flights'') with a power-law distribution $p(\rr,R) \propto A(R,\alpha)r^{-\alpha - 1}, r \rightarrow \infty,  0 < \alpha < 2.$ Besides, a spatially  intermittent magnetic field of the interstellar medium~\cite{Shukurov:2017} results in a higher probability of a long stay of particles in inhomogeneities,  leading to a presence of the so-called ``L\'{e}vy traps''. In the general case, the probability density function 
$q(t,R)$ of time $t$, during which a particle is trapped in the inhomogeneity, also has a power-law behaviour: $q(t,R) \propto B(R,\beta)t^{-\beta - 1},   t \rightarrow \infty,  \beta < 1$.

\item The equation for the density of particles with rigidity $R$ at the location $\rr$ and time $t$, generated in highly inhomogeneous medium by Galactic sources with a distribution density $S(\rr,t,R)$ without energy losses and nuclear interactions can be written as~\cite{Lagutin:2001U,Lagutin:2003}
\begin{equation}~\label{SuperEq}
\frac{\partial N(\rr,t,R)}{\partial t}=-D(R,\alpha,\beta)\mathrm{D}_{0+}^{1-\beta}(-\Delta)^{\alpha/2} N(\rr,t,R)+ S(\rr,t,R).
\end{equation}
Here $\mathrm{D}_{0+}^{1-\beta}$ denotes the Riemann-Liouville fractional derivative~\cite{Samko:1993} and $(-\Delta)^{\alpha/2}$ is the fractional Laplacian (``Riesz operator'')~\cite{Samko:1993}. The anomalous diffusion coefficient is $D(R,\alpha,\beta) \sim A(R,\alpha)/B(R,\beta) = D_0(\alpha,\beta){(R/1~\text{GV})}^{\delta}$. 

In case $\alpha=2$, $\beta = 1$ we obtain Ginzburg-Syrovatsky's normal diffusion equation.

\end{enumerate}

The solution of equation~\eqref{SuperEq} for a point impulse source with emission time $T$ and  
power-law injection spectrum $S(\rr,t,R)=S_{0} R^{-\gamma}\delta(\rr) \Theta(T-t)\Theta(t)$ ($\Theta(\tau)$ is the step function)  has the form~\cite{Lagutin:2003}
\begin{equation}\label{eq:solanomdifeq}
N(\rr,t,R)=\frac{S_0 R^{-\gamma}}{D(R,\alpha,\beta)^{3/\alpha}} 
\int\limits_{\max[0,t-T]}^{t}d\tau \tau^{-3\beta/\alpha}\Psi_3^{(\alpha,\beta)}\left(|\rr|(D(r,\alpha,\beta)\tau^{\beta})^{-1/\alpha}\right),
\end{equation}
where $\Psi_3^{(\alpha,\,\beta)}(\rho)$ is the density of the fractional stable distribution~\cite{Uchaikin:1999a}
\begin{equation*}
  \Psi_3^{(\alpha,\,\beta)}(\rho)=\int\limits_0^\infty{g_3^{(\alpha)}({r\tau^\beta})q_1^{(\beta,1)}(\tau)\tau^{3\beta/\alpha}d\tau}.
\vspace*{-3mm}
\end{equation*}

It should be noted that the spectrum~\eqref{eq:solanomdifeq} has the "knee"~\cite{Lagutin:2001,Lagutin:2017V,Lagutin:2001U,Lagutin:2003}. The spectral index for observed particles $\eta$ at the knee rigidity $R=R_k$ is equal to spectral exponent for particles generated by the source: $\eta|_{R=R_k} = \gamma$. 
One can also find from~\eqref{eq:solanomdifeq} that at $R\ll R_k$ and $R\gg R_k$ we have, respectively:
\begin{equation}\label{eq:index}
\eta|_{R\ll R_k} = \gamma - \delta, \quad \eta|_{R\gg R_k} = \gamma + \delta/\beta.
\end{equation} 

The spectrum of observed primary particles from a steady state  source differs significantly from~\eqref{eq:solanomdifeq}. It has no knee and the spectrum exponent turns out to be equal to the CR spectrum index above the knee~\cite{Lagutin:2001s}:
\begin{equation*}
N_p(\rr,R) \sim R^{-\gamma-\delta/\beta}.
\end{equation*}

The scenario proposed in the paper also  assumes acceleration of secondary particles in Galactic sources. Some mechanisms for the acceleration of these nuclei were discussed in papers~\cite{Berezhko:2003,Blasi:2009,Tomassetti:2012Sec,Berezhko:2014,Cholis:2017}. Since the equilibrium  spectra of secondary CRs are $R^{- \delta}$ times softer than the primary ones, the spectrum of observed secondary particles from a steady state  source in our model can be represented as
\begin{equation*}
N_{s}(\rr,R)\sim R^{-\gamma-\delta/\beta - \delta}.
\end{equation*}

\section{Results}

In the framework of the model described above the observed spectrum of the cosmic ray nucleus $i$ due to all sources of the Galaxy may be presented in the form

\begin{multline}\label{eq:sol}
J(\rr,t,R) = \frac{v}{4\pi}\Biggl[S_{G}R^{-\gamma-\delta/\beta - \varepsilon\delta}+ \frac{S_L R^{-\gamma}}{D(R,\alpha,\beta)^{3/\alpha}}\times\Biggr.\\
\Biggl.\times\sum\limits_{\substack{r_j < 1\;\text{kpc}\\t_j < 10^6\;\text{yr}}} \int\limits_{\max[0,t_j-T]}^{t_j}d\tau	\tau^{-3\beta/\alpha}\Psi_3^{(\alpha,\beta)}\left(|\rr_j|(D(R,\alpha,\beta)\tau^{\beta})^{-1/\alpha}\right)\Biggr].
\end{multline}

It is clear from the physical point of view that the bulk of observed CRs with rigidity $R \sim 1 \div 10^2$   GV forms by numerous distant sources. It means that the CR fluxes in this rigidity region must be described by first term in~\eqref{eq:sol}. The parameter $\varepsilon$ in this term is zero for primary nuclei and is assumed to be equal to one --- for secondary nuclei. The second term defines the spectrum in the high energy region and, as discussed above, has a knee.

In the framework of our approach, we can self-consistently retrieve the main parameters of the diffusion model from experimental data. The key element in the retrieval of these model parameters is the presence of a knee in the spectrum~\eqref{eq:sol}. From equations~\eqref{eq:index}, for example,  we find 
$$\gamma = \eta|_{R\ll R_k} + \delta ,\quad \beta = \frac{\delta}{\eta|_{R\gg R_k} - \eta|_{R\ll R_k} - \delta}.$$
Since  $\eta|_{R\ll R_k}\sim 2.56\div 2.64$, $\eta|_{R\gg R_k} - \eta|_{R\ll R_k}\sim 0.6\div 0.7$~\cite{Bartoli:2015AR,Bartoli:2015}, $\delta\sim 0.27$~\cite{Aguilar:2018Li}, last equations permit to estimate both spectral exponents $\gamma$ and $\beta$:
$$\gamma\sim 2.8\div 2.9,\quad \beta\sim 0.6\div 0.8.$$

We note that a similar steep spectrum of accelerated particles has been observed from the supernova remnants W44~\cite{AbdoW44:2010} and IC 443~\cite{AbdoIC443:2010} with the Fermi Large Area  Telescope, and W49B with H.E.S.S. and Fermi-LAT~\cite{Abdalla:2018}. The value $\gamma$ = 3 has been found in~\cite{Tanaka:2008} for RX J1713.7-3946.  At energies $> 400$ GeV, VERITAS and Fermi-LAT observe gamma-ray emission from Tychos SNR with power-law index $\sim 2.92$~\cite{Archambault:2017}.

To evaluate parameter $\alpha$, general results for particles spectral exponent $\gamma$ obtained in the framework of the diffusive shock acceleration theory extended to the case of anomalous transport with ``L\'{e}vy flights'' and ``L\'{e}vy traps'' have been used (see [Lagutin A.A., 2019, to be published]). It was shown that for $\beta=0.6\div 0.8$ spectral exponent $\gamma=2.8\div 2.9$ corresponds to nondiffusive transport with $\alpha\sim 1.7\div 1.8$.

A set of anomalous diffusion model parameters adopted in this paper is given in Table~\ref{tab:adparams}.
The spatial and temporal coordinates of the local sources are presented in~\cite{Lagutin:2009}. The spherically symmetric force model~\cite{Gleeson:1968} with $\varphi= 600$ MV is used to describe the solar modulation.

\begin{center}
\begin{table}[h]
\centering
\caption{The  anomalous diffusion model parameters}\label{tab:adparams}
\begin{tabular}{@{}l*{15}{l}}
\br
Parameter & Value \\
\mr
$\gamma$ & $2.85$ \\
$\delta$ & $0.27$ \\
$D_0(\alpha,\beta)$ & $ 1.5\cdot 10^{-3}$~pc$^{1.7}$y$^{-0.8}$\\
$\alpha$ & $1.7$\\
$\beta$ & $0.8$ \\
$T$ & $10^4$ y\\
\br
\end{tabular}
\end{table}
\end{center}

\begin{figure}[h]

\includegraphics[width=\textwidth]{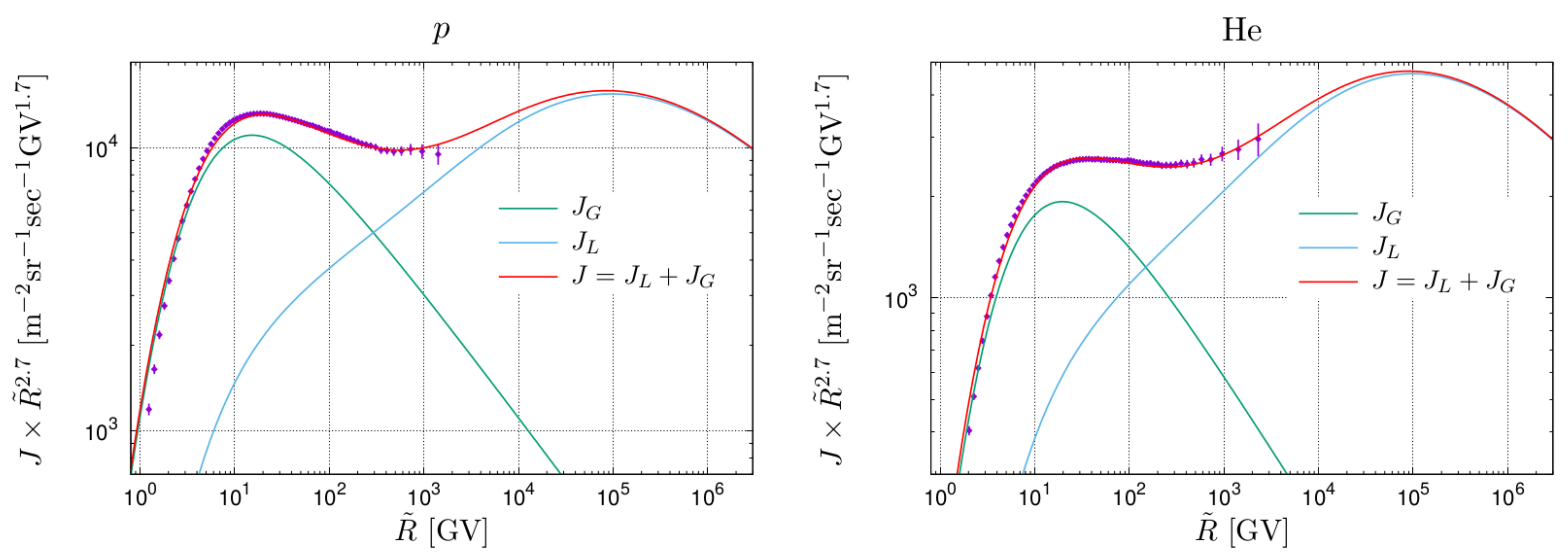}
\includegraphics[width=\textwidth]{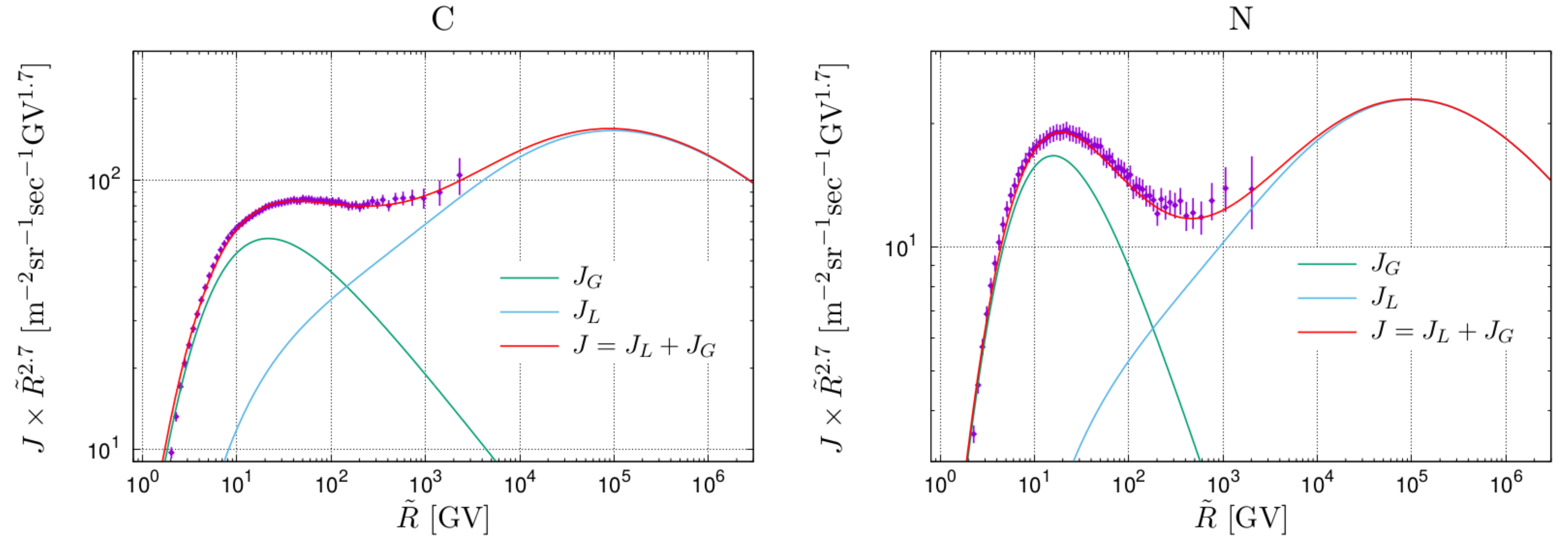}
\includegraphics[width=\textwidth]{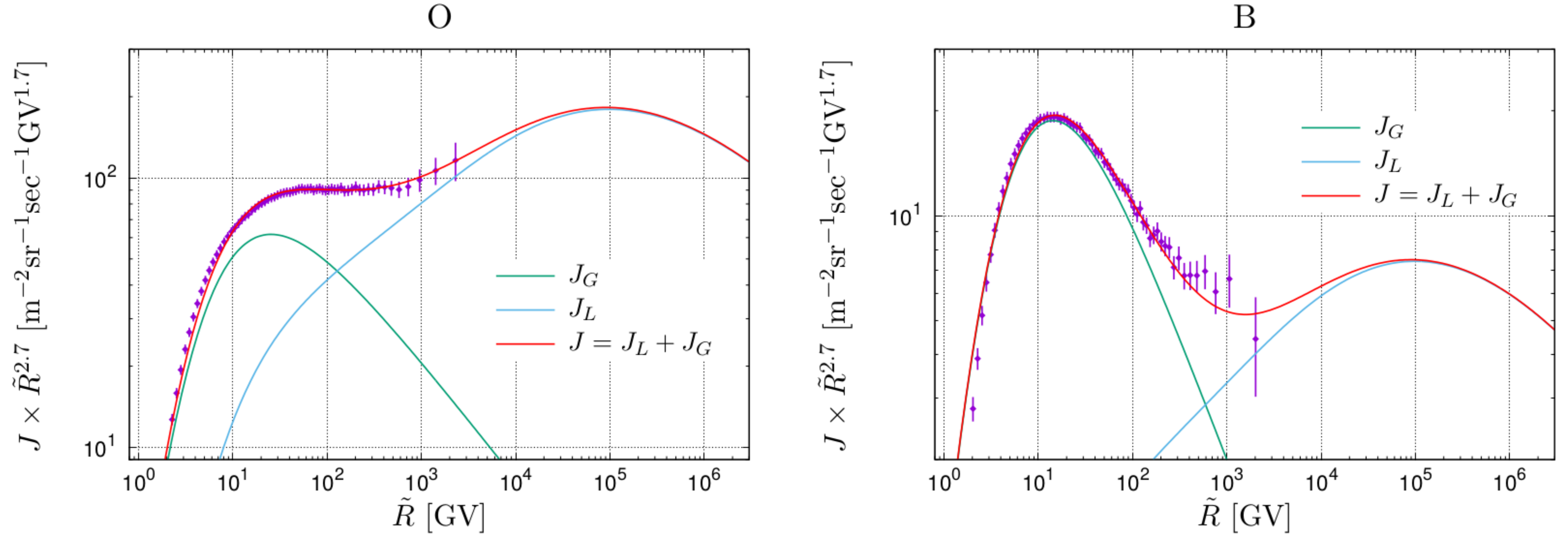}

\caption{Spectra of $p$, He, C, N, O and B multiplied by $\tilde{R}^{2.7}$ as a function of rigidity, obtained in the anomalous diffusion model, compared with the AMS-02 measurements~\cite{Aguilar:2015P,Aguilar:2017CO}}\label{fig:spectra}
\end{figure}

Figure~\ref{fig:spectra} shows the model results of the spectra of nuclei measured by AMS-02 experiment~\cite{Aguilar:2015P,Aguilar:2015,Aguilar:2017CO,Aguilar:2018Li,Aguilar:2018N}.
Points on this figure are the results of AMS-02. In this and the subsequent figures, the data points are placed along the abscissa at $\tilde{R}$ calculated for a flux $f(R) = R^{-2.7}$ with the use of equation (5) from~\cite{Lafferty:1995}.
The solid green lines are the global spectrum component determined by the multiple old distant sources. Solid blue lines are the contribution of the local sources. The spectrum due to all sources of the Galaxy is shown by the solid red line.

\begin{figure}[hp]

\begin{center}
\includegraphics[width=.8\textwidth]{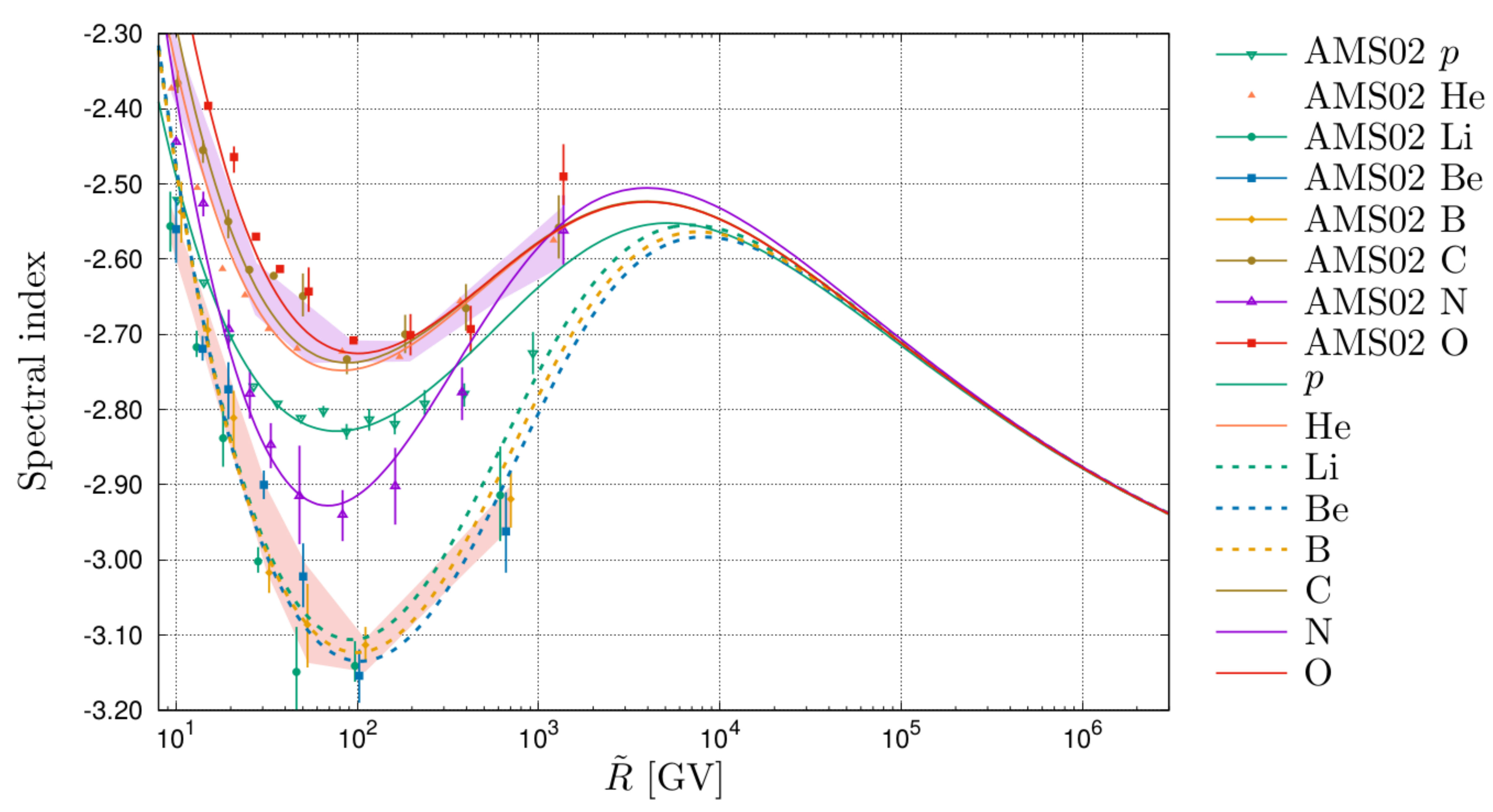}
\end{center}
\caption{The dependence of the $p$, He, C , N, O and Li, Be, B spectral indices on rigidity in the anomalous diffusion model. Data points and the shaded region are from~\cite{Aguilar:2015,Aguilar:2018Li}}\label{fig:spectral-index}

\includegraphics[width=\textwidth]{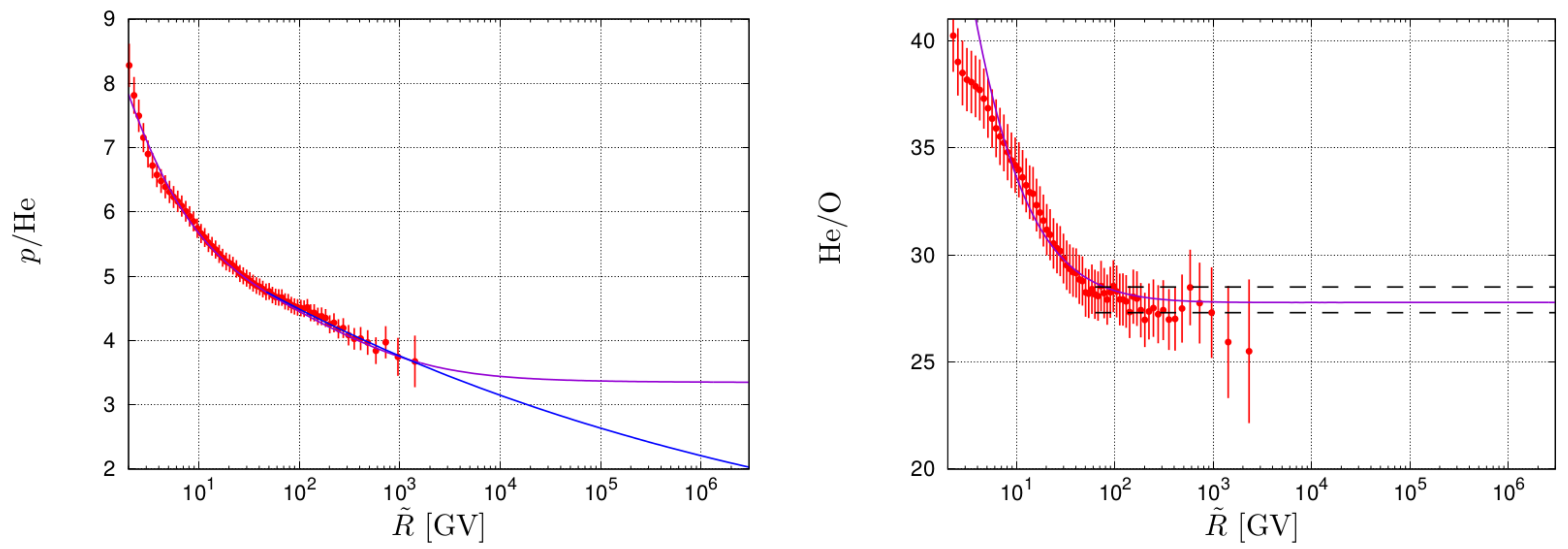}
\includegraphics[width=\textwidth]{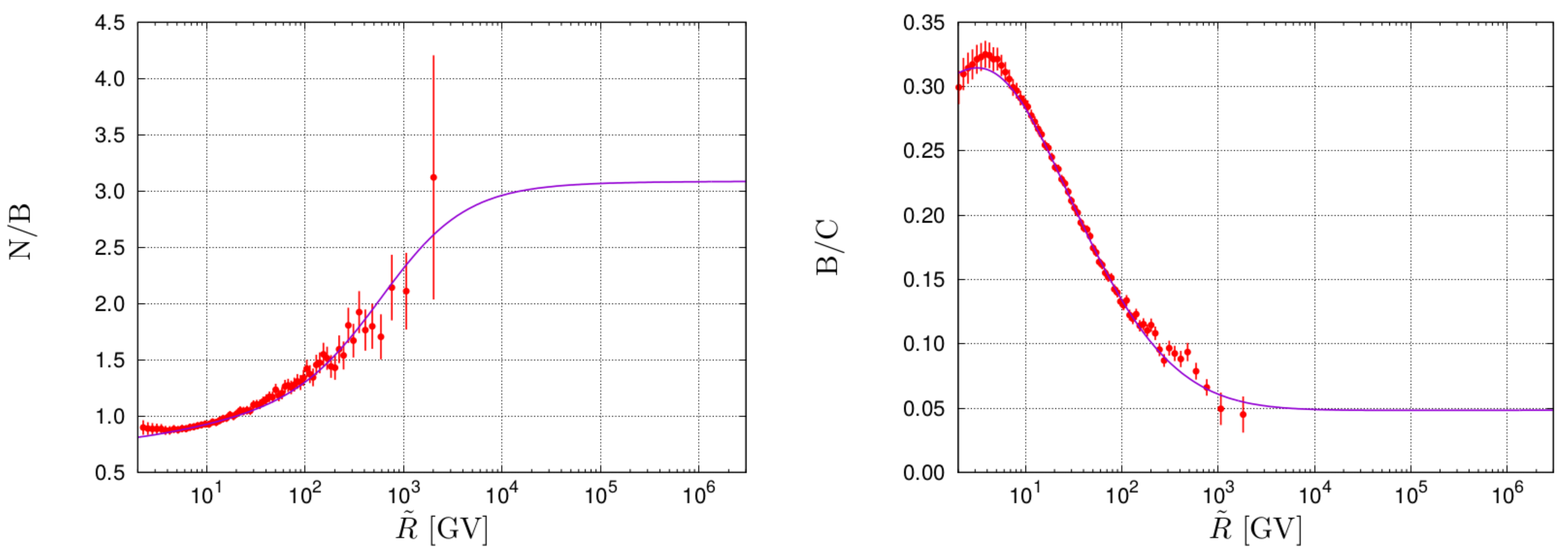}

\caption{The $p$/He, He/O, N/B and B/C flux ratios as functions of rigidity, obtained in the anomalous diffusion model (solid purple lines), compared with the AMS-02 data~\cite{Aguilar:2015,Aguilar:2017CO}. The solid blue curve indicates the fit of a single power law with $\gamma_{p/\text{He}}\approx -0.077$ to the AMS-02 data for $p$/He ratio~\cite{Aguilar:2015}. The dashed black lines indicate the total errors of the fits by constant values of the He/O ratio~\cite{Aguilar:2017CO}}\label{fig:ratios} 
\end{figure}

Figures~\ref{fig:spectral-index} and~\ref{fig:ratios}  show the rigidity dependence of the model spectral index of nuclei measured by AMS-02 and the flux ratios of some nuclei.

\section{Conclusions}

We have presented a novel scenario reproducing the features in the spectra of nuclei measured by AMS-02 experiment~\cite{Aguilar:2015P,Aguilar:2015,Aguilar:2017CO,Aguilar:2018Li,Aguilar:2018N}.
The key elements of this scenario are the following.

\begin{enumerate}
\item All particles with a rigidity $R \lesssim  5 \cdot 10^{7}$ GV, observed on the Earth, are accelerated mainly by galactic sources. The spectrum of these particles in the sources is $J\propto R^{-\gamma}$. 
\item The highly inhomogeneous distribution of matter and magnetic fields in the Galaxy  leads to the anomalous diffusion of CRs.
\item All galactic sources are divided into two groups --- multiple old distant and nearby young ones. This separation is due to a significant difference in the spectra of distant and local sources. A much steeper spectra in GV--TV rigidity region of nuclei  accelerated in distant sources, compared with the spectra from local sources, is determined by steady state mode of particle injection by an ensemble of these distant sources,  as well as by nuclear interactions during their subsequeant propagation in the interstellar medium.
\item Self-consistent technology to retrieve the main parameters of the model from experimental data is used. It is based on the fact that the spectrum of individual  nucleus has a knee.
\end{enumerate}

We demonstrate that the observed changes in the slope   of the cosmic ray spectra caused by the transition from the contribution of multiple distant Galactic sources to the contribution of mainly local ones. We also found that the spectral universality is observed at rigidity $R> 10^5$~GV.
 
The space missions AMS-02, DAMPE and CALET, which currently measure nuclear spectra with high precision in a wide rigidity region, could verify our model predictions in the near future.

\ack

The authors are deeply grateful to the anonymous referee for valuable comments and suggestions, which helped to improve the paper.

This work was supported in part by the Russian Foundation for Basic Research, grant no. 16-02-01103.

\section*{References}
\bibliographystyle{iopart-num}
\bibliography{ID292-Lagutin}

\end{document}